\pdfoutput=1
\documentclass[sigconf,nonacm,screen]{acmart}
\AtBeginDocument{%
  }

\setcopyright{none}
\usepackage{pdflscape}
\usepackage{tabularx}
\usepackage{booktabs}
\usepackage{array}
\usepackage{enumitem}
\usepackage{subcaption}
\usepackage{soul}

\begin{document}

\title[\textsc{Waypoint}]{\textsc{Waypoint}: Collaborative Planning and Authorising with Coding Agents}

\author{Ludwig Felder}
\email{ludwig.felder@tum.de}
\affiliation{%
  \institution{Technical University of Munich}
  \city{Munich}
  \country{Germany}}

\author{Leon Cena}
\email{leon.cena@tum.de}
\affiliation{%
  \institution{Technical University of Munich}
  \city{Munich}
  \country{Germany}}

\author{Chunyang Chen}
\email{chunyang.chen@tum.de}
\affiliation{%
  \institution{Technical University of Munich}
  \city{Munich}
  \country{Germany}}

\renewcommand{\shortauthors}{Felder et al.}

\begin{abstract}
When delegating tasks to autonomous coding agents, developers increasingly rely on plan modes to review intended changes before execution. However, because current agents generate these plans as flat conversational text, developers struggle to pinpoint flawed assumptions, attach localised feedback, or verify if a chatbot's revision addresses their concerns. As agents tackle more complex, long-running tasks, this intermediate step becomes a critical bottleneck. Flawed, unquestioned assumptions during planning may lead to incorrect implementations and lost trust. To make pre-execution oversight collaborative, we built \textsc{Waypoint}, a system informed by a formative study ($N=10$) that replaces chat outputs with structured, editable plans. A controlled, within-subjects study ($N=16$) compared this system against native plan modes. \textsc{Waypoint} enabled users to address software behaviour and verification inside plans more frequently and locally. Users reported greater control reviewing intended work while spending more time reviewing. Our study explores the design of collaborative planning and authorisation. 
\end{abstract}



\begin{teaserfigure}
    \centering
    \includegraphics[width=1\linewidth]{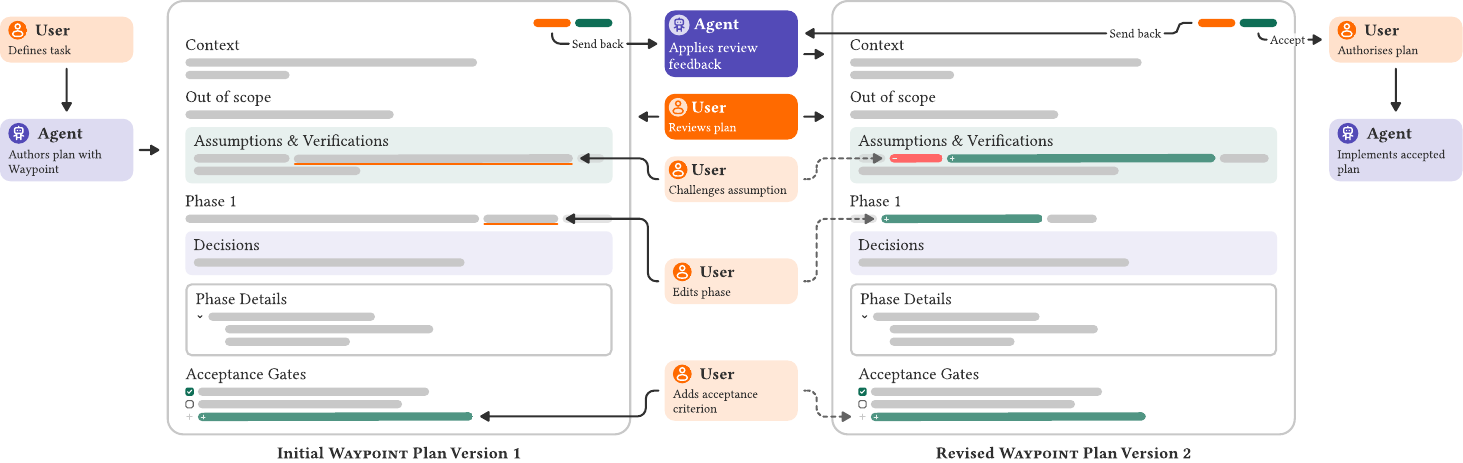}
    \caption{\textsc{Waypoint}'s collaborative plan-review workflow. The agent first generates plan version 1. The developer reviews the plan and proposes revisions through the central interactions. They can challenge assumptions, edit phase content, or add acceptance criteria. The developer then sends the plan back to the agent to integrate the feedback into version 2 of the plan. This review and revision loop can repeat until the developer authorises the plan for implementation.}
    \Description{Schematic showing how review actions connect an initial Waypoint plan on the left to a revised plan on the right. Between the plans, three developer actions point to an assumption, phase text, and an acceptance checklist. Dotted arrows connect these locations to corresponding revisions in version 2, where additions and deletions are marked. The plans share the same section structure, with placeholder lines representing text. Send back arrows from both versions lead to the agent's feedback-integration step. A separate Accept arrow from version 2 leads to developer authorisation and then agent implementation.}
    \label{fig:header}
\end{teaserfigure}

\maketitle

\section{Introduction}

Generative AI tools have become an essential part of today's software engineering practices. Since their introduction, they shifted from conversational assistance toward delegated autonomous implementation. While coding assistants generated answers or completions to questions, coding agents can inspect a repository, use tools and modify multiple files with less continuous developer input~\citep{chenCodeMeMe2026,huangProfessionalSoftwareDevelopers2025}. Today, agents fix bugs, implement features and commit code~\cite{robbesAgenticMuchAdoption2026}, with more than 450k pull requests across 61k+ repositories on GitHub~\cite{liAIDevStudyingAICoding2026}. At the same time, the capabilities of the underlying LLMs, along with their surrounding software infrastructure and control layer, are improving. Thereby enabling agents to implement increasingly complex and long-running tasks~\cite{desaiSWEMarathonCanAgents2026,deng_swe-bench_2025}. As developers delegate these tasks to agents, planning is emerging as an intermediate step during implementation.

Coding agents like Claude Code\footnote{\url{https://code.claude.com/docs/en/permission-modes}}, Codex\footnote{\url{https://learn.chatgpt.com/guides/best-practices}} or Cursor\footnote{\url{https://cursor.com/docs/agent/plan-mode}} expose this stage through dedicated \emph{plan mode}. 
Instead of directly implementing code changes, the agent first explores the codebase to understand the problem. For example, if a requested feature requires a new user interface component, the agent reads related files to identify where the component belongs and to examine how neighbouring components are structured. It searches for their dependencies and runs terminal commands to trace how data flows between the interface and the backend. Subsequently, the agent drafts an implementation plan, specifying which files need to be modified and how. The original request states the desired change, while the plan proposes how the agent intends to implement it. The developer can review and revise this proposal before implementation begins. The resulting code then records what the agent actually changed.

A plan can therefore be understood as a detailed instruction prompt derived from the original user request. The plan serves two actors: first, the executing agent, and second, the human engineer who checks the plan. 
For the agent, planning can improve the generated code at the function and repository levels~\cite{ICLR2025_071a637d,bairiCodePlanRepositoryLevelCoding2024}. A developer's request that is misunderstood by the agent and subsequently translated into an incorrect implementation results in the error being detected only at the last stage. The effect, however, depends on plan quality, as a subpar plan can lead to worse performance than no plan at all~\citep{liuPlanActionHow2026}.
For developers, the same plan externalises intended actions before edits to the codebase are made. This thereby creates an opportunity to align the agent's understanding of the task with the request. Assumptions can be questioned, requirements corrected, or implementation choices changed or questioned. Developers identified planning as one form of agentic oversight~\cite{dhanorkarHumanOversightAgentic2026}. This collaboration between the human and agent, however, can be brittle if users misplace trust in the agent's plan, or cannot understand how the plan will be translated into action~\cite{hePlanThenExecuteEmpiricalStudy2025}.

Current agents already allow developers to inspect plans, request revisions, and, in some tools, to edit directly. Research systems extend this interaction by making task decompositions editable, placing approval before coding, or using shared plans to distribute work and revise them during execution~\citep{kazemitabaarImprovingSteeringVerification2024a,takerngsaksiriHumanIntheLoopSoftwareDevelopment2025,fengCocoaCoPlanningCoExecution2026a}. These approaches show that plan review is not a simple accept-or-reject gate. However, coding-agent plans can contain many proposed changes, assumptions, and implementation details without indicating which parts require the developer's attention. Developers must then locate relevant information, explain through conversation which part concerns them, and determine whether the agent's revision addresses their feedback. A developer's review should remain traceable through revision, authorisation, and implementation. This leaves open the question of how to support developers and coding agents in jointly inspecting, discussing, and changing a plan before coding begins. We therefore ask: \textbf{How can a system support developers in reviewing and revising an implementation plan before implementation by an agent?}

We investigate this question through formative interviews, system design, and a user and system evaluation. Interviews with 10 practitioners yielded five design goals and informed \textsc{Waypoint}, a system for collaboratively reviewing and revising coding-agent plans. Named for a checkpoint along a route, \textsc{Waypoint} combines a structured plan, a review interface, localised feedback, and an agent revision loop before implementation. We compared \textsc{Waypoint} with native planning in Claude Code and Codex in a counterbalanced within-subjects study with 16 participants, then autonomously implemented all 32 accepted plans. Using \textsc{Waypoint}, participants focused more of their interventions on required software behaviour and verification, and less on plan presentation. They also reported greater control and readiness to accept the agent's intended work. Review and revision took longer with \textsc{Waypoint}. We discuss three implications for the design of coding-agent systems that support collaborative plan review.

Concretely, our work makes the following contributions:
\begin{enumerate}[label=\textbf{\arabic*.},leftmargin=*]
    \item \textbf{Design requirements for collaborative plan review.}
    Formative interviews with 10 software practitioners characterise how developers review coding-agent plans and inform five design goals for supporting review and revision before implementation.
    \item \textbf{\textsc{Waypoint}, a system for collaborative plan review and revision.}    
    \textsc{Waypoint} combines a structured plan format, a review interface, and agent workflows that help developers inspect assumptions and decisions, provide localised feedback, and inspect the revised plan before authorising implementation.    
    \item \textbf{Findings on plan review and authorisation.}
    A counterbalanced within-subjects study with 16 developers compares review practices, reported experience, and review time across native and \textsc{Waypoint} workflows. Subsequent implementation of all 32 accepted plans examines adherence to participant-introduced commitments.
\end{enumerate}

Together, these contributions position collaborative plan review as an important design direction for software engineering with increasingly autonomous agents. Although we study coding, the same interaction problem may arise wherever people need to inspect and revise an agent's proposed work before it acts.

\section{Background and Related Work}\label{sec:rw}

\subsection{Coding Agents and Plan-First Workflows}\label{sec:rw-agents}

AI coding assistants initially functioned as copilots, offering support through autocomplete and conversation while developers directed and integrated their output~\citep{chenCodeMeMe2026}. Studies reported productivity gains for this interaction model~\citep{pengImpactAIDeveloper2023b}. Developers nevertheless spent considerable effort verifying and repairing the generated code~\citep{fanWhenHelpHurts2026}. Coding agents extend this model. An agent combines a language model with software that lets it navigate repositories, modify files, execute tools, and pursue longer-running tasks~\citep{yangSWEagentAgentComputerInterfaces2024,wangOpenHandsOpenPlatform2025}, commonly described as a ``harness''~\citep{robbesAgenticMuchAdoption2026}. Advances in language-model reasoning~\citep{weiChainofthoughtPromptingElicits2022}, tool use~\citep{yaoReActSynergizingReasoning2023}, and self-revision~\citep{madaanSelfRefineIterativeRefinement2023,shinnReflexionLanguageAgents2023} allow this autonomy. Agent-authored changes now appear at a substantial scale across public repositories~\citep{liAIDevStudyingAICoding2026,robbesAgenticMuchAdoption2026}. Developers control these agents by scoping tasks, making design decisions, and reviewing their output~\citep{huangProfessionalSoftwareDevelopers2025}. Experienced developers preserve control through detailed delegation and strategic oversight, whereas novices oscillate between over-reliance and avoidance~\citep{fengJuniorSeniorAllocating2026}. Greater autonomy can improve task correctness, but it also risks leaving developers with a poorer understanding of the resulting changes~\citep{chenCodeMeMe2026}. As coding becomes more delegated, maintaining understanding and control therefore becomes an increasingly important and challenging part for software engineers.

Planning has emerged as one control point in this workflow. Repository-level planning methods derive a chain of dependent edits and execute it step by step~\citep{bairiCodePlanRepositoryLevelCoding2024}. Agents expose explicit planning modes or specification phases, including Claude Code\footnote{\url{https://code.claude.com/docs/en/overview}}, Codex\footnote{\url{https://learn.chatgpt.com/guides/best-practices}}, and OpenCode\footnote{\url{https://open-code.ai/en/docs/modes}}. Specification-driven workflows such as GitHub Spec Kit\footnote{\url{https://github.com/github/spec-kit}} and Kiro\footnote{\url{https://kiro.dev}} split planning into a sequence of interrelated documents (specification, technical plan, task list), each reviewed before the next is generated. A study by~\citet{liuPlanActionHow2026} showed across 16{,}991 programming-agent trajectories, appropriate plans improved issue resolution, whereas subpar plans performed worse than providing no explicit plan, and agents frequently deviated from instructed plans. Plans are consequently neither merely explanatory output nor guarantees of faithful execution. Our work examines how this increasingly consequential intermediate artefact can support meaningful human review before implementation is delegated.

\subsection{Plans as Coordination and Review Artefacts}\label{sec:rw-plans}

Plans have long been understood as resources for situated action rather than scripts that fully determine behaviour~\citep{1212951,suchmanResponseVeraSimons1993,rooksbyCanPlansSituated2014,barriramPlansSituatedAction,bardramWhyPlanDoesnt2010}. Even when goals or procedures are available in advance, their concrete implementation must be worked out and may be reformulated as circumstances unfold~\citep{rooksbyCanPlansSituated2014}. Plans can therefore orient action without fully specifying or guaranteeing their accomplishment.

Computer-supported cooperative work research similarly characterises formal artefacts as coordination mechanisms that reduce articulation work and function as both maps and defeasible scripts~\citep{schmidtCoordinationMechanismsConceptual,schmidtMapsScriptsStatus1999}. Design-rationale systems externalise issues, alternatives, and criteria~\citep{conklinGlBISHypertextTool,maclean2020questions}. Such systems often failed in practice, illustrating Grudin's observation that collaborative tools fail when those who do the additional work are not those who benefit~\citep{grudinWhyCSCWApplications1988}: the person recording rationale bears a cost whose benefit accrues to later readers. Generative models shift this cost. An agent can draft a rationale, detect insufficient explanations, and infer missing reasoning for a person to verify rather than author~\citep{sonClearFairyCapturingCreative2026}. Reviewing generated content, however, demands that readers monitor their own understanding and calibrate their confidence in the output, a metacognitive burden that explainability and structure can partly shift to the system~\citep{tankelevitchMetacognitiveDemandsOpportunities2024a}. Anchored annotations, versioning, and displayed edit-history provide established mechanisms for such structure, supporting localised feedback and iterative negotiation over changing artefacts~\citep{brushRobustAnnotationPositioning2001,zytoSuccessfulClassroomDeployment2012,keryVarioliteSupportingExploratory2017,viegasStudyingCooperationConflict2004}.

Software engineering offers a particularly relevant precedent for reviewing intended work. Design documents, requirements traces, and architecture decisions all record proposed changes and their rationale~\citep{parnasRationalDesignProcess1986,gotelAnalysisRequirementsTraceability1994,buchgeherUsingArchitectureDecision2023}. Modern code review provides a lightweight, tool-mediated, and iterative process for scrutinising such changes~\citep{sadowskiModernCodeReview2018,rigbyConvergentContemporarySoftware2013}. Its dominant activity is understanding rather than merely finding defects~\citep{bacchelliExpectationsOutcomesChallenges2013}, with reviewers seeking information about alternatives and evidence that their understanding is correct~\citep{pascarellaInformationNeedsContemporary2018}. Review feedback also creates downstream coordination work, motivating systems that couple comments with actionable edits~\citep{froemmgenResolvingCodeReview2024}. Finally, contracts and acceptance criteria connect intended behaviour to assessable outcomes~\citep{meyerApplyingDesignContract1992,solisStudyCharacteristicsBehaviour2011,984017}, while the oracle problem marks where correctness still requires human judgement~\citep{barrOracleProblemSoftware2015}. \textsc{Waypoint} brings these principles upstream by treating the implementation plan as a persistent artefact for localised review, revision, and assessment before code is produced.

\subsection{Interactive Planning and Agent Oversight}\label{sec:rw-oversight}

Prior HCI systems establish plans as useful surfaces for human--agent interaction. Editable steps and assumptions can improve steering and verification relative to conversational prompting~\citep{kazemitabaarImprovingSteeringVerification2024a}, although reviewing a plausible plan before execution can also increase misplaced trust~\citep{hePlanThenExecuteEmpiricalStudy2025}. In software engineering, HULA requires engineers to approve a file-level coding plan before implementation, with 82\% of successfully generated plans approved in deployment~\citep{takerngsaksiriHumanIntheLoopSoftwareDevelopment2025}. Furthermore, Cocoa presents a plan as a shared artefact through which users modify steps, allocate work between themselves and an agent, inspect intermediate outputs, and re-plan as execution unfolds~\citep{fengCocoaCoPlanningCoExecution2026a}. These systems demonstrate that exposing a plan creates opportunities for intervention, but approval or editability alone does not establish how carefully a plan was reviewed or understood.

Interviews with experienced developers describe agent oversight through a priori configuration, collaborative planning, real-time monitoring, and post hoc review, with monitoring rarely performed in practice~\citep{dhanorkarHumanOversightAgentic2026}. Participants used co-planning to refine intent before execution and sometimes treated plans as proxies for agents' actual work, assuming adherence to reduce code-review effort. Existing interfaces support action guards, proactive pauses, and steering during execution~\citep{mozannarMagenticUIHumanintheloopAgentic2025,pengMoraeProactivelyPausing2025,eppersonInteractiveDebuggingSteering2025}. Initiative can also lie with the assistant, which decides from the programmer's context when to offer a suggestion; such proactivity raised task completion, but its acceptance depended on the timing and integration of suggestions~\citep{chenNeedHelpDesigning2025a}. \textsc{Waypoint} focuses specifically on the pre-execution boundary that arises when a developer intends to delegate implementation rather than participate continuously in it. Whereas Cocoa uses the plan to distribute agency across interleaved planning and execution, \textsc{Waypoint} uses the plan as an artefact for a developer to review and authorise subsequent autonomous work. A step list tells the reviewer what the agent intends to do, not which steps rest on guesses about the codebase or where the agent is likely to be wrong. Generated output should make its uncertainties visible, show the sources behind it, and add friction at the points where users decide to rely on it~\citep{weiszDesignPrinciplesGenerative2024a}. Language models, however, ask fewer clarifying questions than humans and frequently leave underspecification unresolved~\citep{shaikhNavigatingRiftsHumanLLM,vijayvargiyaAmbigSWEInteractiveAgents2026}. \textsc{Waypoint} combines these strands at the pre-execution boundary, addressing calls for greater steerability and verifiability in coding-agent research~\citep{wangHumansAreMissing2026}. Plans should expose assumptions, provenance, decisions, and acceptance criteria for localised review and revision. After authorisation and implementation, a persistent plan with evidence of execution allows users to compare it to their commitments.

\subsection{Review, Verification, and Calibrated Reliance}\label{sec:rw-reliance}

Effective oversight depends not only on whether users can intervene, but on whether their reliance reflects an automation's actual trustworthiness~\citep{leeTrustAutomationDesigning}. As more functions are delegated~\citep{parasuramanModelTypesLevels2000}, people are left with difficult residual tasks such as monitoring behaviour, diagnosing failures, and intervening under exceptional conditions~\citep{BAINBRIDGE1983775}.
Studies show how far self-assessment can drift from reality. Experienced open-source developers took longer on their own issues with AI tools yet believed the tools had sped them up~\citep{beckerMeasuringImpactEarly2025}, and participants writing code with an AI assistant produced more security vulnerabilities while rating their code as more secure~\citep{perryUsersWriteMore2023}. Developers' trust in code-generation tools shifts with the task's stakes and complexity, yet the tools provide few signals for validating output or expressing intent~\citep{wangInvestigatingDesigningTrust2024}. Verification is also costly, accounting for 22.4\% of session time in one study~\citep{mozannarReadingLinesModeling2024b}, while reviews of agent-generated code can require more interaction than reviews of human-written code~\citep{zhongHumanAISynergyAgentic2026}. Longitudinal evidence further associates repeated exposure to agent-authored pull requests with increasing approval and declining inline-comment activity~\citep{yuHabituationGate2026}. Human supervision alone, therefore, does not guarantee effective supervision.

Verification behaviour is nevertheless sensitive to interface design. Analysts establish confidence by consulting multiple supporting artefacts~\citep{guHowAnalystsUnderstand2024a}. Reducing engagement costs can lower overreliance~\citep{vasconcelosExplanationsCanReduce2023}, and uncertainty presentations and live execution change how programmers inspect generated code~\citep{vasconcelosGenerationProbabilitiesAre2024,ferdowsiValidatingAIGeneratedCode2024a}. The form of evidence also matters. Explanations increase acceptance of AI recommendations regardless of their correctness~\citep{bansalDoesWholeExceed2021a,kimFosteringAppropriateReliance2025}, whereas sources can improve calibration~\citep{kimFosteringAppropriateReliance2025}. Plans raise the same problem. Plausible plans invite misplaced trust~\citep{hePlanThenExecuteEmpiricalStudy2025}, the plans people prefer are not reliably the ones that help them execute a task~\citep{balepurGoodPlanHard}, and subpar plans degrade agent performance~\citep{liuPlanActionHow2026}. Whereas these studies assess preference, execution success, or agent adherence separately, our work combines how developers rate a plan, how they revise it, and whether the commitments are reflected in the plan and its implementation.

\section{Formative Study}\label{sec:formative}

Prior work shows how plans can support steering, approval, or co-planning with agents across different applications \citep{kazemitabaarImprovingSteeringVerification2024a, takerngsaksiriHumanIntheLoopSoftwareDevelopment2025, fengCocoaCoPlanningCoExecution2026a}. Yet little is known about how software practitioners use agent-generated plans in their everyday work. We conducted a formative study aiming to understand when plans are requested and how they are reviewed and accepted. To ground the design of \textsc{Waypoint}, we derived design goals from the findings of this study.

\subsection{Participants, Procedure, and Analysis}

We recruited 10 software practitioners (I1-I10) through the authors' professional networks. Nine practitioners worked across several large software companies, and one participant was a student with prior software engineering experience. Participants ranged from interns to professionals with more than 10 years of experience and worked on distributed systems, frontend libraries, machine-learning pipelines, developer frameworks, or customer-facing prototypes. Agentic tools used across the sample included GitHub Copilot and the Copilot CLI, Cursor, Claude Code, and Kiro.

All interviews were conducted in individual video calls and lasted approximately 30 to 45 minutes. We asked when and why participants requested plans, how they inspected, and when they revised or rejected them. Further, we scoped how they judge plan quality and the challenges they face during planning, plan revision, and implementation. With informed consent, we recorded and transcribed the interviews and redacted project-specific details. We analysed the transcripts using inductive thematic analysis \citep{braunUsingThematicAnalysis2006}. Both authors independently open-coded each transcript. We then compared and consolidated the codes, checked the themes against the complete corpus, and resolved differences through discussion.

\subsection{Planning Practices}\label{sec:planning-process}

Participants calibrated planning effort to a task's scope, uncertainty, and consequences. Small changes often went directly to an agent because planning could slow them down (I5, I6). Larger features, unfamiliar systems, cross-team dependencies, and sensitive components received more planning (I4, I7, I9). I9 reasoned that planning becomes more valuable as feature size increases because \textit{``uncertainties and inaccuracies increase''}. Participants also described the cost of skipping both planning and revising plans. I10 described early experiences of working without plan that produced \textit{``a huge changeset that I cannot review''}. I8 valued detailed plans, but faced \textit{``quite a lot of work''} when the agent failed to apply revisions. The most common plan artefact was a single Markdown file refined in chat, although I4, I5, and I10 also described multi-document workflows. We focus on the review and revision loop across single-plan documents rather than on comparing the specific planning processes.

\subsection{Challenges in Plan Review and Use}\label{sec:findings}

We organise our findings into five design problems, ordered across the plan workflow. They move from reviewer attention and agent understanding through revision and implementation to preserving rationale.

\paragraph{Plan review is shallow and can become rubberstamping (F1)}
Six participants described skimming, time-capping, or passing through plan review (I1 to I3, I6, I7, I9). I2 spent \textit{``less than five minutes''} on a plan but inspected the resulting code line by line. I6 would read at most one page of the document before accepting. Five participants (I1 to I4, I6) had accepted plans they did not fully understand, often because a later implementation was reversible or they expected to inspect the result instead. This occasionally deferred costs, with I6 reporting that reading a skipped plan section would have saved 30 minutes during deployment. I9 also found that repeated agent confirmations trained them to press \textit{``Enter, Enter, Enter, yes, yes, yes''} instead of discussing the plan. 
Together, these reports describe three forms of selective review: limited reading, acceptance despite incomplete understanding, and routine confirmation. Hence, agent-generated plans should direct limited attention to consequential decisions.

\paragraph{A plan's validity depends on inputs it does not expose (F2)}
Participants often checked whether the agent understood their request rather than whether the agent's proposal would work. This was especially difficult in unfamiliar code (I1, I6). I4 described a plan that overlooked a network constraint documented deep within another team's wiki, rendering the proposed dependency impossible. I7 noted that other constraints existed only \textit{``in someone's head''} and could not be found in code or documentation. I3 identified their main blind spot as \textit{``whether the right context was set''}. These accounts suggest that polished plans can hide uncertainty about their premises. Reviewable plans should expose their assumptions, supporting evidence, and research boundaries so that practitioners can identify missing organisational knowledge.

\paragraph{Feedback and navigation are indirect and coarse (F3)}
Participants revised plans via chat, but flat conversations made it difficult to address the content. I2 encountered multiple numbered lists that each restarted at 1, while I6 struggled with hundreds of terminal lines and wanted to cross-reference two documents. I8 gave the model one revision attempt before editing the plan manually and found cross-cutting changes particularly likely to fail. I10 had already improvised a solution by placing comments directly in plan documents and asking the agent to \textit{``check my comments''}. These practices suggest that plan review needs addressable content. Reviewers should be able to attach local or cross-cutting feedback directly to the plan and retain those anchors across revisions.

\paragraph{An accepted plan does not bind its implementation (F4)}
Participants distinguished agreement with a plan from confidence in its execution. I5 cautioned that \textit{``having a very tight plan in place does not mean the output will match your expectations''}. I8 had seen an agent explain requirements correctly and then omit them during implementation. Participants instead relied on small-increment implementation and testing (I5, I8, I10). I10 wanted plans to include unit tests, expected logs, and instructions for manually checking the application. At the same time, requirements could change during implementation (I6), and exploratory work remained partly uncertain (I10). A plan should therefore make commitments verifiable and deviations visible without assuming that implementation can follow it mechanically~\citep{barriramPlansSituatedAction,bardramWhyPlanDoesnt2010}.

\paragraph{Plans can carry understanding, but their rationale is fragile (F5)}
Some participants delegated most code authoring to agents (I6, I7, I10). I10 reported barely reviewing the completed code because \textit{``I already saw it in the markdown [plan] before''}, using the plan to identify the core classes worth inspecting. Yet the rationale produced during planning could disappear with the session. I9 argued that a change was generated \textit{``must not be lost''}, and I8 valued returning to conversations weeks or months later. Participants proposed keeping plans with the repository (I5, I10), including using pull requests for line-level plan review (I5). These accounts suggest that plans can become records of understanding for changes developers did not author line by line. Such records should preserve the accepted plan, its negotiation, and evidence from execution.

\subsection{Design Goals}\label{sec:goals}

Together, the findings characterise reviewable plans as attention-efficient (F1), grounded (F2), addressable (F3), accountable to implementation (F4), and persistent (F5). We translated these dimensions into five design goals for \textsc{Waypoint}:

\begin{enumerate}[
    label=\textbf{DG\arabic*.},
    ref=DG\arabic*,
    leftmargin=*
]
    \item\label{dg:attention}
    \textbf{Budget reviewer attention (F1).}
    Foreground consequential or convention-deviating decisions with their rationale while allowing implementer-oriented detail to recede.
    
    \item\label{dg:assumptions}
    \textbf{Surface and qualify assumptions (F2).}
    Distinguish premises supported by repository evidence, external claims, or planned checks from those that the agent merely asserts.

    \item\label{dg:verification}
    \textbf{Make commitments verifiable and deviations legible (F4, F5).}
    Attach executable checks where possible, explicitly mark human-judged criteria, and record evidence of execution and any departures from the plan.

    \item\label{dg:feedback}
    \textbf{Anchor feedback to plan content (F3).}
    Support comments on specific sections, cross-cutting requests, and direct edits while preserving anchors across revisions.

    \item\label{dg:persistence}
    \textbf{Persist the plan and its negotiation (F5, F4).}
    Keep the plan, rationale, revision history, and execution record with the repository so they survive the originating session.

\end{enumerate}

These goals span plan generation and repository research (A1), the plan artefact (A2), its review interface (A3), and execution (A4). \textsc{Waypoint} directly designs A1 to A3 and guides the agent in A4 to implement the accepted plan and produce evidence records (DG3 and DG5, Figure~\ref{fig:flow}).

\section{Waypoint Interface and System}\label{sec:system}

\begin{figure}[t]
    \centering
    \includegraphics[width=1\linewidth]{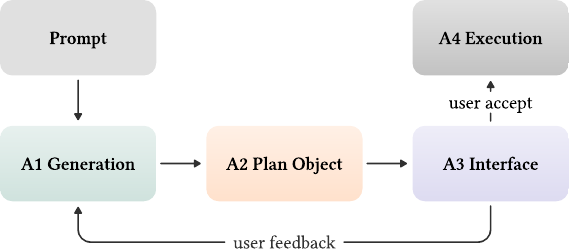}
    \caption{\textsc{Waypoint} restructures the boundary before autonomous implementation. The agent researches the repository and authors a persistent plan (A1 and A2). The developer reviews and revises that artefact in the interface (A3). Only explicit acceptance authorises the agent to begin implementation (A4). Feedback returns to generation without starting execution.}
    \Description{Five-box flowchart with a feedback loop below the main path. The prompt at the upper left points down to A1, Generation, then right to A2, Plan Object, and A3, Interface. From the interface, an arrow labelled user feedback loops back to generation. A separate arrow labelled user accept points upward to A4, Execution, at the upper right.}\label{fig:flow}
\end{figure}

Guided by our design goals, we present \textsc{Waypoint}, a system for reviewing and authorising coding-agent plans before autonomous implementation. Its central artefact is a repository-local plan that moves through agent generation, developer review and revision, explicit acceptance, and later execution. \textsc{Waypoint} combines agent instructions, a structured plan format, a plan-centred interface, and companion workflows that incorporate feedback and implement accepted plans. Throughout this section, we connect these components to DG1-DG5 in Section~\ref{sec:goals}.

We abstract the \textsc{Waypoint} system into four areas of action (Figure~\ref{fig:flow}). Following a user request, the agent researches the codebase and generates a plan (A1) as a structured artefact (A2). The generated plan is then reviewed and revised in a dedicated interface (A3), and the agent incorporates human feedback into a revised plan. Only an explicit acceptance action authorises the plan for autonomous implementation (A4). We describe the plan artefact first because generation, review, and execution all operate on this shared representation.

\subsection{Reviewable Plan Artefact}\label{sec:system-artifact}

\begin{figure*}[tbp]
    \centering
    \includegraphics[width=0.7\linewidth]{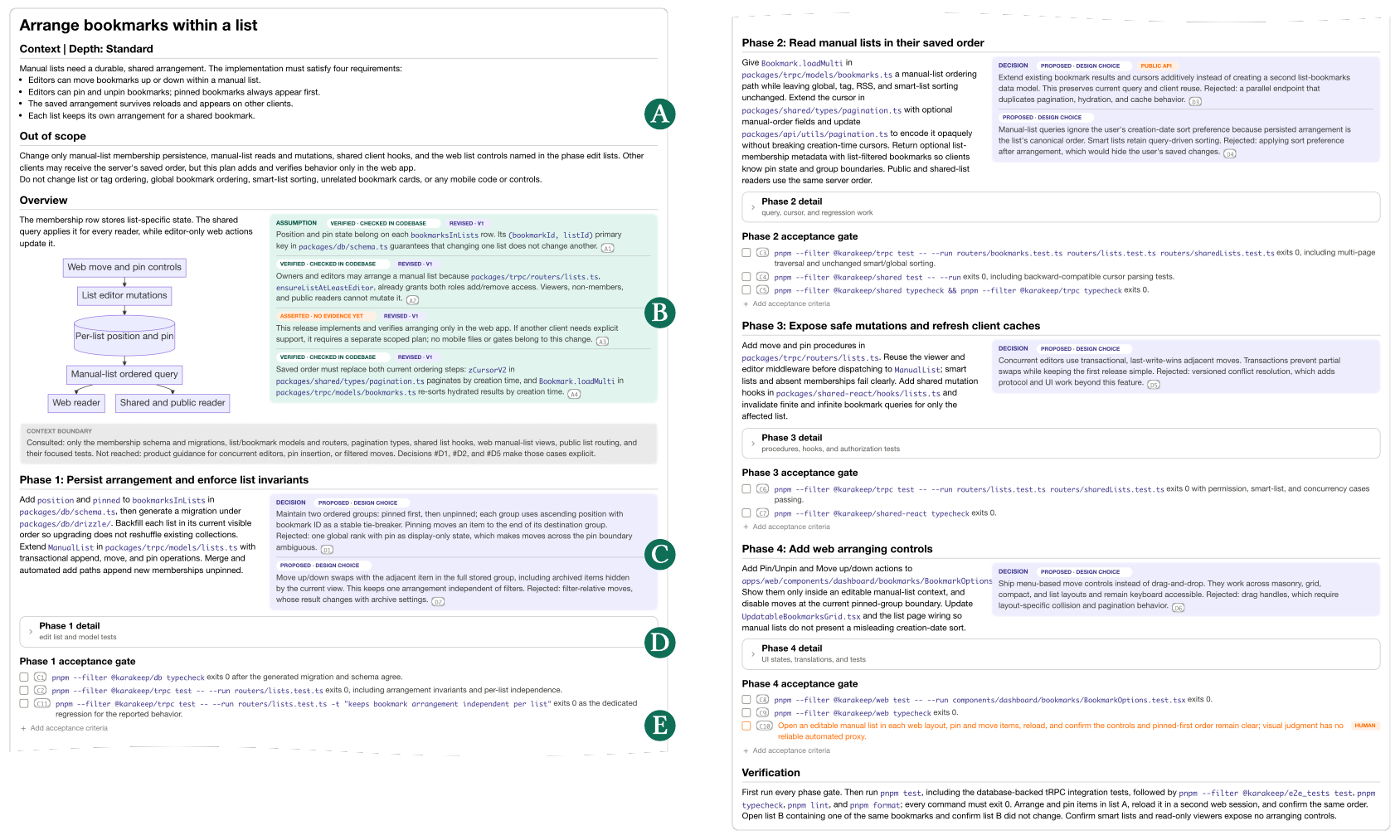}
    \caption{Complete \textsc{Waypoint} plan rendered for review. Annotations highlight: (A) scope and research boundaries (Section~\ref{sec:system-scope}), (B) evidence-tagged assumptions (Section~\ref{sec:system-assumption}), (C) decisions and rejected alternatives (Section~\ref{sec:system-decisions}), (D) collapsed implementer detail (Section~\ref{sec:system-disclosure}), and (E) acceptance gates that distinguish agent-checkable tests from criteria requiring human decision (Section~\ref{sec:system-commitments}).}
    \Description{Annotated screenshot of a plan for arranging bookmarks within a list, split into two side-by-side segments. Read the left segment from task context through phase 1, then the right segment through phases 2 to 4 and final verification. Phase summaries sit beside decision callouts, with collapsed detail rows and acceptance checklists below. The overview includes a diagram linking web controls, list mutations, stored positions and pin states, and the ordered query used by web, shared, and public readers. Assumption badges distinguish claims checked in the codebase from an assertion without evidence. Decision callouts show rationales and rejected alternatives. Checklist entries pair identifiers with checks and expected outcomes, with one visual check explicitly marked Human.}\label{fig:interface}
\end{figure*}

The plan artefact is a Markdown file stored in the repository. It therefore survives the agent session and remains available to humans, and later agent runs (DG5). \textsc{Waypoint} extends Markdown with constructs that expose the information needed for review while retaining the detail required for implementation.

\subsubsection{Scope and Research Boundary}\label{sec:system-scope}

The plan opens with a \emph{Context} section stating the expected result and an \emph{Out of scope} section recording what the plan excludes (Figure~\ref{fig:interface}A). A \emph{context boundary} then discloses which code, documentation, and external systems the agent inspected and which sources it could not reach. These fields allow reviewers to judge the proposal against an explicit research boundary rather than infer its completeness from a polished plan.

\subsubsection{Assumption}\label{sec:system-assumption}

\emph{Assumption} sections expose consequential premises (\ref{dg:assumptions}, Figure~\ref{fig:interface}B). Each assumption distinguishes evidence observed in the repository, a claim from a named external source, a premise linked to a planned gate check, or an unsupported assertion. The repository evidence names the relevant file and symbol. A premise that is not supported by repository evidence must either link to a gate that will test it or state the consequence if it is false. This distinction keeps evidence already observed separate from evidence that execution is expected to be obtained later.

\subsubsection{Decisions}\label{sec:system-decisions}

Phases foreground novel or convention-deviating choices as \emph{decision} callouts (\ref{dg:attention}, Figure~\ref{fig:interface}C). Each callout describes the selected approach, its rationale, and rejected alternatives. The agent may also flag a decision as deviating from the request, the existing code, or a general convention, and may flag stakes such as a public interface or security boundary. These flags direct the reviewer's attention to review agent assumptions.

\subsubsection{Verification Gates}\label{sec:system-commitments}

Each phase ends with a gate based on binary criteria (DG3, Figure~\ref{fig:interface}E). The gates map intentions to verification evidence. Agent-checkable criteria name the command or test that decides them and the expected result. Human-judged criteria are reserved for outcomes for which no reliable automated proxy exists, such as visual layout or interaction quality, and are explicitly marked. \textsc{Waypoint} instructs the agent to seek developer input on decisions that can be settled during planning. During execution, the agent can't close these human gates. A final \emph{Verification} section connects the phase-level gates to an end-to-end check. 

\subsubsection{Progressive Disclosure for Two Readers}\label{sec:system-disclosure}

The same file serves readers with asymmetric needs (Figure~\ref{fig:interface}D). The developer needs enough information to decide whether the proposed change is appropriate. The implementing agent needs the full edit list, commands, and code-level detail. \textsc{Waypoint} keeps scope, assumptions, decisions, phase intent, and gates visible, while placing implementer-oriented material inside expandable blocks (\ref{dg:attention}). The authoring instructions also prefer linked pointers to repository files and symbols over copied code. Detail is therefore relocated rather than removed. A reviewer can open it on demand, while a later agent receives the complete artefact. Plan-wide identifiers preserve assumptions, decisions, and criteria across revisions. Each review round seals the plan text, comments, and compiled review request as a read-only version, retaining both the accepted plan and the negotiation through which it became acceptable (DG5, Figure~\ref{fig:review-round}).

\subsection{Generating Plans for Review}\label{sec:system-generation}

Developers start the workflow by invoking the repository-installed \emph{waypoint} skill with a coding task. Skills are instruction files loaded on demand by the coding agent\footnote{\url{https://agentskills.io}}, thereby not requiring modifying the underlying agent. The skill places the agent in a read-only planning phase. The agent may inspect the repository, run checks and raise clarifying questions, but the plan is the only project file it may write. Table~\ref{tab:a1-generation} summarises the skill rules that shape the resulting plan.

\begin{table}[t]
    \caption{The \emph{waypoint} skill instructs agent actions derived from the design goals. These behaviours define how the agent authors a plan for review.}\label{tab:a1-generation}
    \Description{Two-column table pairing four agent behaviours with their effects on a generated plan. Read each row from the behaviour on the left to the corresponding authoring actions on the right. The rows cover consequential choices, research boundaries, checkable commitments, and reviewability. Design-goal references accompany each behaviour.}
    \small
    \setlength{\tabcolsep}{4pt}
    \renewcommand{\arraystretch}{1.12}
    \begin{tabularx}{\columnwidth}{>{\raggedright\arraybackslash}p{0.34\columnwidth} >{\raggedright\arraybackslash}X}
        \toprule
        \textbf{Agent behaviour} & \textbf{Effect on the generated plan} \\
        \midrule
        Foreground consequential choices (\ref{dg:attention}) & Mark novel or convention-deviating decisions and state their rationale and rejected alternative. \\
        Establish the research boundary (\ref{dg:assumptions}) & Inspect relevant code and dependencies, record the source of important claims, and disclose what could not be reached. \\
        Define checkable commitments (\ref{dg:verification}) & Express completion as binary criteria with a command and expected result, or mark the criterion as requiring human review. \\
        Edit the draft for reviewability (\ref{dg:assumptions}, \ref{dg:attention}) & Run an extraction pass for hidden assumptions, then move agent-only detail into collapses and replace copied code with pointers. \\
        \bottomrule
    \end{tabularx}
\end{table}

The agent inspects the relevant code, identifies integration boundaries, records what it cannot inspect, designs the implementation, and asks the developer about choices that cannot be resolved from the repository.

Plan generation ends with two passes by the agent over the finished plan. An \emph{extraction pass} rereads the completed draft to surface premises that were used but not stated (\ref{dg:assumptions}). A \emph{compression pass} rereads the visible layer from the reviewer's perspective. It retains information required to judge scope, premises, decisions, and commitments, moves implementation detail into expandable blocks, and replaces copied repository content with linked file and symbol references (\ref{dg:attention}).

\subsection{Reviewing, Revising, and Authorising Plans}\label{sec:system-interface}

The \textsc{Waypoint} viewer is a webview editor pane within the IDE. It monitors the repository plan directory and renders each incremental write, so reviewers can begin reading while the agent is authoring the plan. Assumptions and decisions appear as addressable callouts, acceptance gates as interactive checklists, and implementer detail as expandable sections. References to repository files open in an adjacent editor, allowing the reviewer to compare a claim with the code on which it rests in one click.

\begin{figure}[tbp]
    \centering
    \begin{subfigure}[t]{.48\linewidth}
        \centering
        \includegraphics[width=\linewidth]{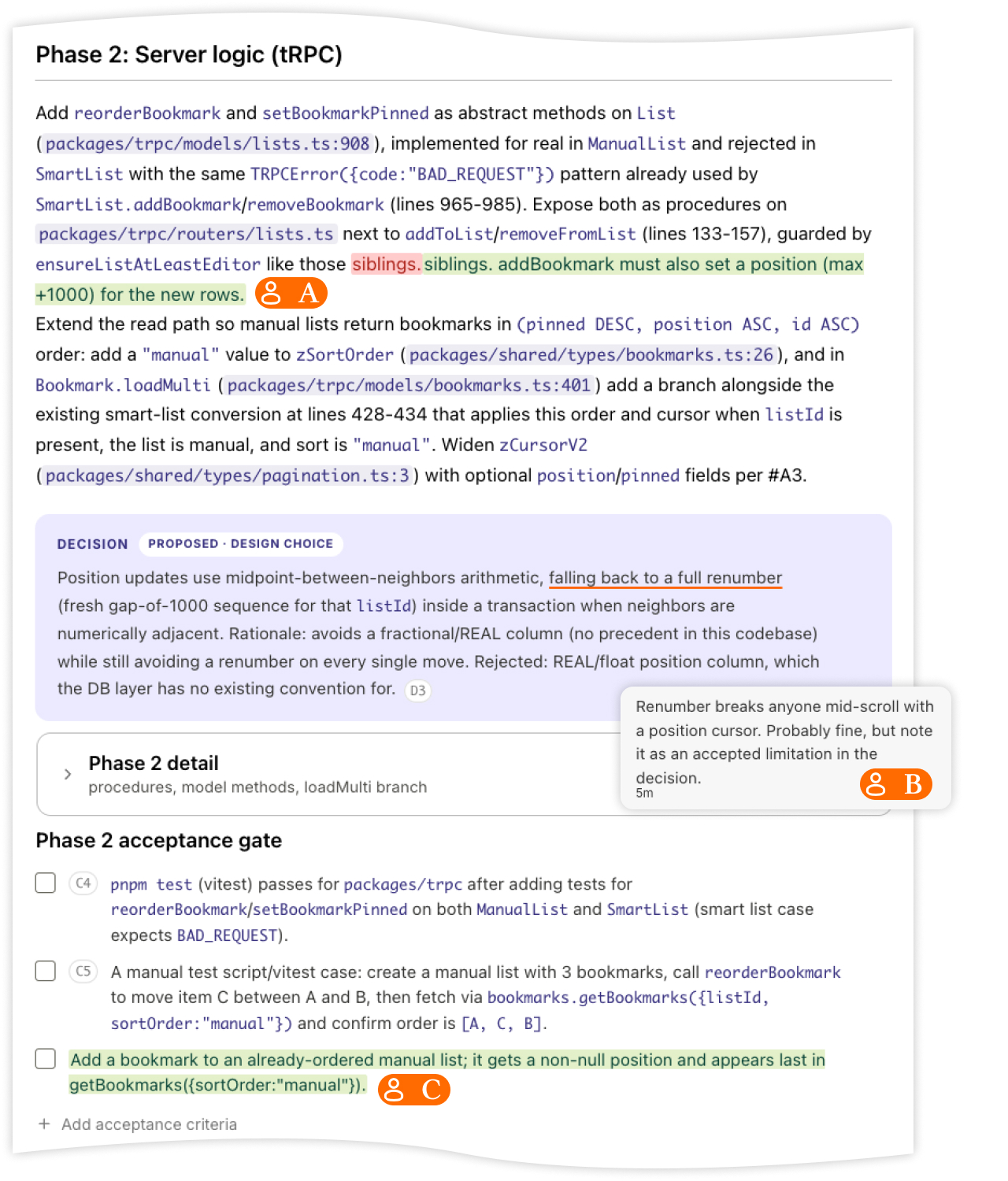}
        \caption{Reviewer actions: a direct edit (A), an anchored comment (B), and an added criterion (C).}\label{fig:interactions}
    \end{subfigure}
    \hfill
    \begin{subfigure}[t]{.48\linewidth}
        \centering
        \includegraphics[width=\linewidth]{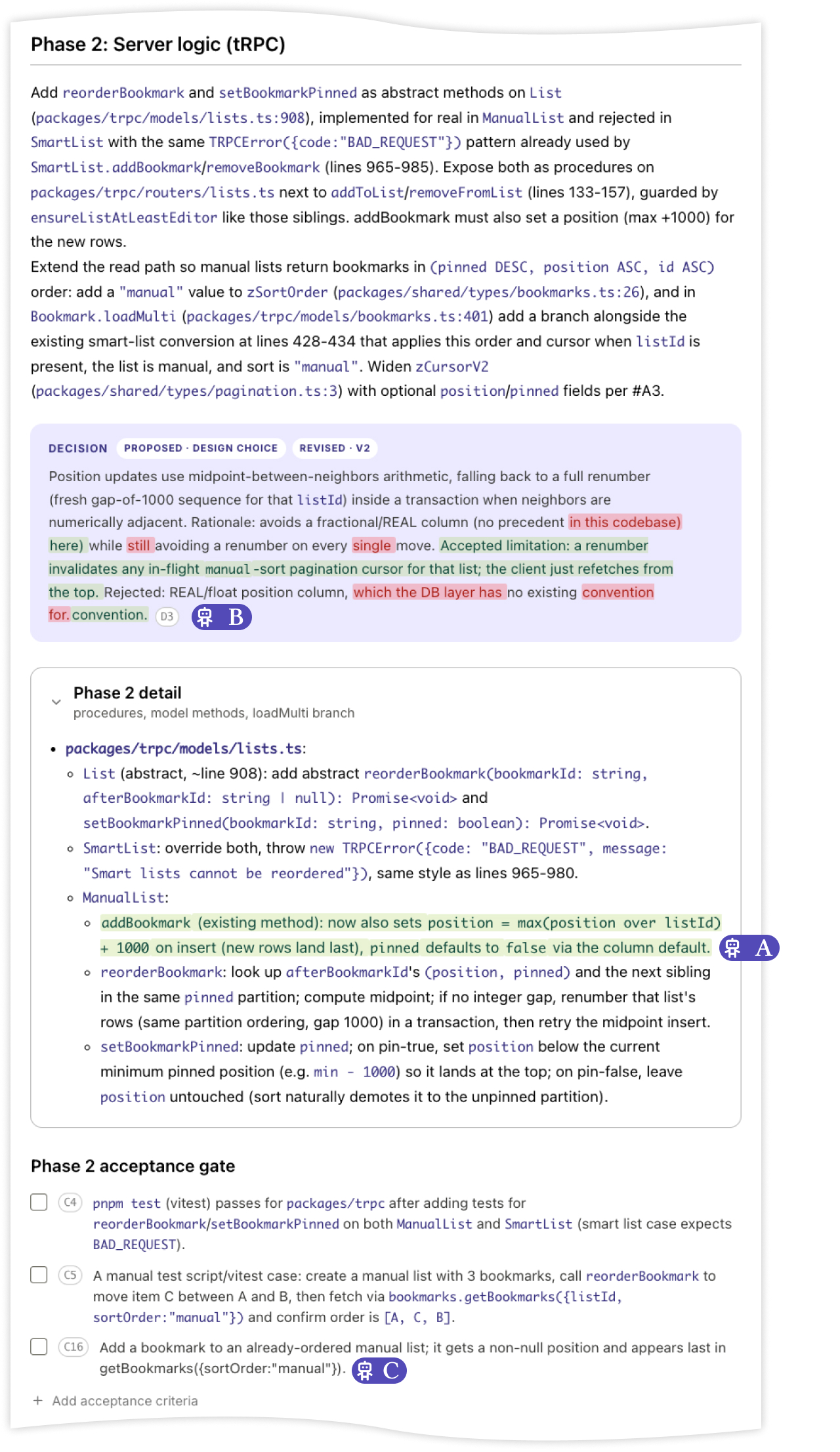}
        \caption{The revision preserves the comment's intent and records its change in version 2 (B), propagates the direct edit (A), and retains the added criterion under a stable ID (C).}\label{fig:revision}
    \end{subfigure}
    \caption{One review round in \textsc{Waypoint}: the reviewer acts on the plan (a), and the agent applies the review to produce the revised plan (b).}
    \Description{Paired screenshots show a review of the server-logic phase on the left and the agent's revision on the right. In A, the developer adds that new bookmarks receive a position 1,000 above the current maximum. The revision carries this rule into the expanded implementation details. In B, an anchored comment asks the agent to acknowledge that renumbering disrupts a reader's current pagination position. The revised decision records this limitation and says the client fetches again from the top, with highlighted text changes and a version 2 badge. In C, the developer adds a check that a new bookmark appears last in an already ordered list. The revision retains this criterion with identifier C16.}\label{fig:review-round}
\end{figure}

Reviewers can act on the plan in four ways. They can select any content and attach a local comment, or mark the comment as cross-cutting when it requires a whole-plan revision. Users can reject an assumption or decision directly, which otherwise are assumed accepted, avoiding a sequence of mandatory confirmations that could reproduce rubberstamping (F1). Reviewers can also edit the plan directly. Waypoint saves each change to the plan file and highlights the added and removed text (Figure~\ref{fig:interactions}, panel~A). Finally, reviewers can add or remove acceptance criteria (Figure~\ref{fig:interactions}, panel~C). Clicking \emph{Send back} compiles the entire round into a single review request and seals the plan, its comments, and that request as a read-only version. The companion \emph{waypoint-apply} skill integrates the review into the existing plan in a new version (Figure~\ref{fig:revision}).

\subsection{Execution After Authorisation}\label{sec:system-execution}

Clicking \emph{Accept plan} seals the reviewed plan and passes it to the companion \emph{waypoint-implement} skill for autonomous execution. During implementation, the skill instructs the agent to verify the results by passing the gates. A record of evidence is attached to each agent-checkable criterion. These records include the checked state and the successful command, output, or retained log run. A failed check requires an explicit override reason, while a human gate remains for human review. These implementation records preserve the decisions during implementation (\ref{dg:persistence}).

\subsection{Implementation}\label{sec:system-implementation}

\textsc{Waypoint} is implemented as a Visual Studio Code extension in TypeScript, with the viewer built as a webview using Svelte\footnote{\url{https://svelte.dev}}. The system keeps all state in the user's repository and uses no backend. Plans, active comments, review requests, edit baselines, and sealed versions remain local files under a dedicated repository directory. A file watcher connects the agent writes to the live view, which is also how incremental plan revisions appear while the agent is writing. Markdown is rendered with marked\footnote{\url{https://marked.js.org}} and sanitised before display. Code is highlighted with highlight.js\footnote{\url{https://highlightjs.org}}, and diagrams are rendered with Mermaid\footnote{\url{https://mermaid.js.org}}. Custom fenced constructs represent assumptions, decisions, boundaries, collapses, checklists, and execution records. The extension invokes the main \emph{waypoint} skill and its \emph{waypoint-apply} and \emph{waypoint-implement} companions through the agent, providing an integrated terminal. The extension is published on the Visual Studio Code Marketplace as \emph{Waypoint Plan Reviewer}, and its source code is available at \url{https://github.com/ludwigfelder/waypoint_extension} (Appendix~\ref{sec}).

For our evaluation, the extension logs all interactions to a local file, including edits, review rounds, and per-section dwell times by viewport share. We additionally record the agents' LLM traffic at the wire level with mitmproxy\footnote{\url{https://www.mitmproxy.org}}, recording the request and response streams of agents without modifying them.

\section{User and System Evaluation}\label{sec:study}

We evaluated \textsc{Waypoint} in two parts. First, we conducted a counterbalanced within-subjects study in which developers planned and reviewed one implementation using \textsc{Waypoint} and another using their assigned agent's native plan mode. Participants stopped after accepting a plan and did not implement the change themselves. They did not inspect the later implementation. Secondly, we then provided the accepted plans to coding agents for autonomous implementation in fresh environments to assess the resulting implementations. Our design thereby examines both the experience of reviewing a plan and explores traceability from accepted commitments to implementation.
Specifically, our evaluation aimed to address the following research questions:

\begin{enumerate}[label=\textbf{RQ\arabic*.},ref=RQ\arabic*,leftmargin=*]
    \item\label{rq:review}\textbf{Review Process:} How does plan review differ between the \textsc{Waypoint} and native workflows as developers inspect, question, and revise agent-generated implementation plans?
    \item\label{rq:understanding}\textbf{Understanding and Authorisation :} How do participants report understanding and readiness to authorise across the \textsc{Waypoint} and native workflows?
    \item\label{rq:accountability}\textbf{Implementation Accountability:} How do commitments accepted during plan review carry into autonomous implementation, and were they followed?
\end{enumerate}

\begin{figure}[t]
    \centering
    \includegraphics[width=\linewidth]{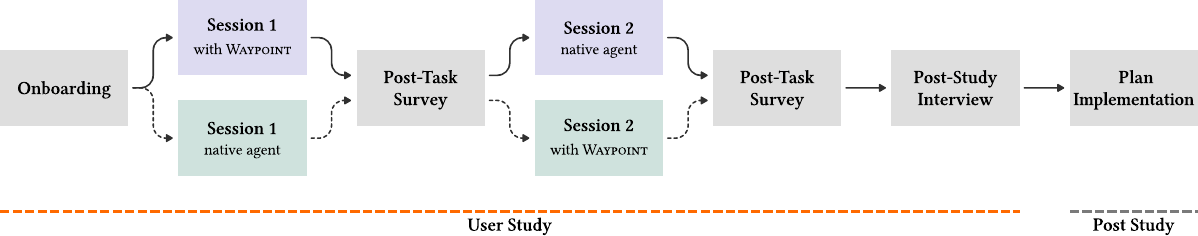}
    \caption{Study overview. After onboarding, participants completed two counterbalanced planning sessions with \textsc{Waypoint} and native plan mode. Each session used a different feature task and ended with a questionnaire and comprehension quiz. Participants then answered comparison questions and completed a post-study interview. After the study, we autonomously implemented the accepted plans through their corresponding workflows.}
    \Description{Left-to-right flowchart showing two condition orders. After onboarding, the upper path uses purple boxes and solid arrows for Waypoint in session 1 and the native agent in session 2. The lower path uses green boxes and dashed arrows for the reverse order. The native agent label refers to native plan mode. Both paths pass through a post-task survey after each session, then a shared post-study interview and plan implementation. An orange dashed line beneath onboarding through the interview marks the user study. A separate grey dashed line beneath plan implementation marks the post-study stage.}
    \label{fig:study_flow}
\end{figure}

\subsection{Participants}

We recruited sixteen participants through mailing lists and professional social media. Participants were screened for professional software engineering experience, regular use of coding agents, familiarity with relevant technologies such as TypeScript, React, and SQL, and no prior familiarity with the study codebase. Five were junior developers, two were mid-level developers, two were senior developers, one was a tech lead, four were students, and two were researchers. Two participants reported less than one year of professional software development experience; five reported one to three years; seven reported three to five years; and two reported five to ten years. 

Fourteen used AI coding assistants daily, and two used them weekly. Two participants reported not using a dedicated plan mode 
; four used them rarely; two sometimes; seven often; and one always. Participants were compensated €30. Our study protocol was reviewed and approved by our institution's ethics board. All participants provided informed consent before taking part. 

\subsection{Study Design and Conditions}

We used a within-subjects design to compare \textsc{Waypoint} with each agent's native plan mode (Figure~\ref{fig:study_flow}). Each participant completed both feature tasks and used each condition once. The eight permutations crossed task order (Task~1 or Task~2 first), condition order (\textsc{Waypoint} or native plan mode first), and assigned agent (Claude Code or Codex). The agent was a between-subjects factor. Eight participants used Claude Code with \texttt{anthropic/claude-sonnet-5}, and eight used the Codex CLI with \texttt{gpt-5.6-sol}. Each participant used the same agent and underlying model in both sessions.

\subsection{Tasks}

Participants planned two feature tickets for Karakeep\footnote{\url{https://karakeep.app}}, an open-source bookmark manager. We selected Karakeep as a representative production web application with a frontend, backend, database, background workers, and search system. None of the participants had prior familiarity with the repository. To mitigate the risk of the agent just reproducing an existing solution from its training data, both tickets came from unresolved issues. We selected the tasks for their clear requirements and cross-component scope. At the same time, they remained small enough for participants to review a complete plan within the session. \emph{Task 1: Recoverable Delete} required soft deletion, restoration, permanent deletion, and exclusion of deleted content from views and search. Automatic cleanup was optional. Trash support for other entities and a user-configurable cleanup interval were excluded. \emph{Task 2: Manual List Ordering} required persistent ordering and pinning within each list. Other forms of ordering were excluded. Both tasks required changes across the database, API, and interface. Appendix~\ref{sec:study-tasks} provides the complete requirements and exclusions.

\subsection{Procedure}

Sessions were conducted individually, with a target duration of approximately 70 minutes.
After providing consent, participants connected VS Code to a prepared virtual machine containing the repository, agents, \textsc{Waypoint}, and logging setup. They watched an introduction to Karakeep and, before the \textsc{Waypoint} condition, a tutorial of its features. For each task, participants worked with the assigned agent until they believed the plan could be implemented autonomously and reliably. We allowed participants to decide for themselves when to authorise their plans, so some sessions ran beyond that target. During each task, they could inspect the repository and the running application throughout.

After each task, participants completed a post-task questionnaire including a three-question comprehension quiz. They answered the quiz from memory without reopening the plan, repository, or transcript. After the \textsc{Waypoint} condition, they also rated its different features. The study concluded with three comparison questions about the two planning modes and a semi-structured interview.

\subsection{Measures and Analysis}\label{sec:analysis}

We logged complete agent trajectories in both conditions and derived planning and revision times, plan submissions, participant prompts, model calls, token usage, and cost. \textsc{Waypoint} additionally recorded comments, direct edits, assumptions and decision verdicts, checklist changes, review rounds, and viewport-based dwell time. Because native plan modes exposed no equivalent interface events, we use \textsc{Waypoint} feature use and dwell measures descriptively rather than as condition comparisons. For the review analysis, we counted each independently answerable request, assigned a purpose to it, and checked whether it appeared in the accepted plan. We then traced accepted requests concerning behaviour, implementation choices, or verification into the implementation. Appendix~\ref{sec:coding-audit} provides the full codebook inclusion and exclusion rules.

After each condition, participants completed NASA-TLX~\citep{hart1988development}, six items adapted from the Trust in Automation questionnaire~\citep{korber2018theoretical}, and study-specific measures of plan quality, review and steering, self-assessed understanding, and readiness to authorise. These ratings pertained to the entire planning workflow. Two comprehension questions concerned the repository's structure and mechanisms and were scored against a common ground-truth answer key. The third concerned how a specific requirement was addressed and was scored against the participant's accepted plan.

We used two-sided Wilcoxon signed-rank tests for paired comparisons. We analysed the NASA-TLX dimensions, adapted trust statements, and study-specific statements individually. The questionnaire blocks organise related questions but do not define aggregate scales. Questionnaire item analyses used all available complete participant pairs, so denominators vary where participants selected ``No Answer''. We used the forced-choice comparisons and interview responses descriptively to contextualise the questionnaire and interaction results.

\subsection{Autonomous Implementation and Assessment}

Each participant produced one final native plan and one accepted \textsc{Waypoint} plan. We autonomously implemented all 32 plans through their corresponding workflows using the authoring model, a clean checkout, an isolated container, and otherwise matched settings. No human intervened during implementation. We recorded model trajectories, repository diffs, tests, time, tokens, and cost. Human-only and unavailable end-to-end criteria were recorded as bypassed rather than credited as agent verification. We also ran 16 no-plan implementations as a secondary  reference.

We assessed ticket-required behaviour using frozen capability-discovering suites with 15 tests for Task~1 and 13 for Task~2. The suites covered database and tRPC behaviour and relevant query and authorisation paths, but not rendered interfaces, search workers, or background workers. We constructed them after the runs to accommodate heterogeneous APIs. We report type checking, unexpected regressions, agent-authored tests, time, and cost separately. For \textsc{Waypoint}, we also report whether the agent marked acceptance gates as complete and attached evidence records.

To assess implementation adherence, we retained participant requests that concerned required behaviour, implementation choices, or verification and appeared in the accepted plan. We traced each of these commitments into the files the agent changed or created, its recorded actions, and the test results. We classified each commitment as followed or not followed, or as not assessable when the retained artefacts could not establish an outcome. We excluded human-only gates, requests that changed only phrasing or organisation, and criteria written by the agent rather than the participant. Appendix~\ref{sec:coding-audit} documents the procedure and category definitions in full. We report condition-specific counts descriptively because the two workflows produced commitment pools that differed in size and composition.

\section{Results}\label{sec:results}

We compare how participants used \textsc{Waypoint} and native plan mode for reviewing plans and authorising implementation. We first examine how participants reviewed and revised plans (RQ1), then their experience and understanding before authorisation (RQ2), and finally how agents implemented the accepted plans (RQ3).

\subsection{RQ1: Review Process}\label{sec:results-rq1}

\begin{figure*}[!t]
    \centering
    \includegraphics[width=\linewidth,height=.82\textheight,keepaspectratio]{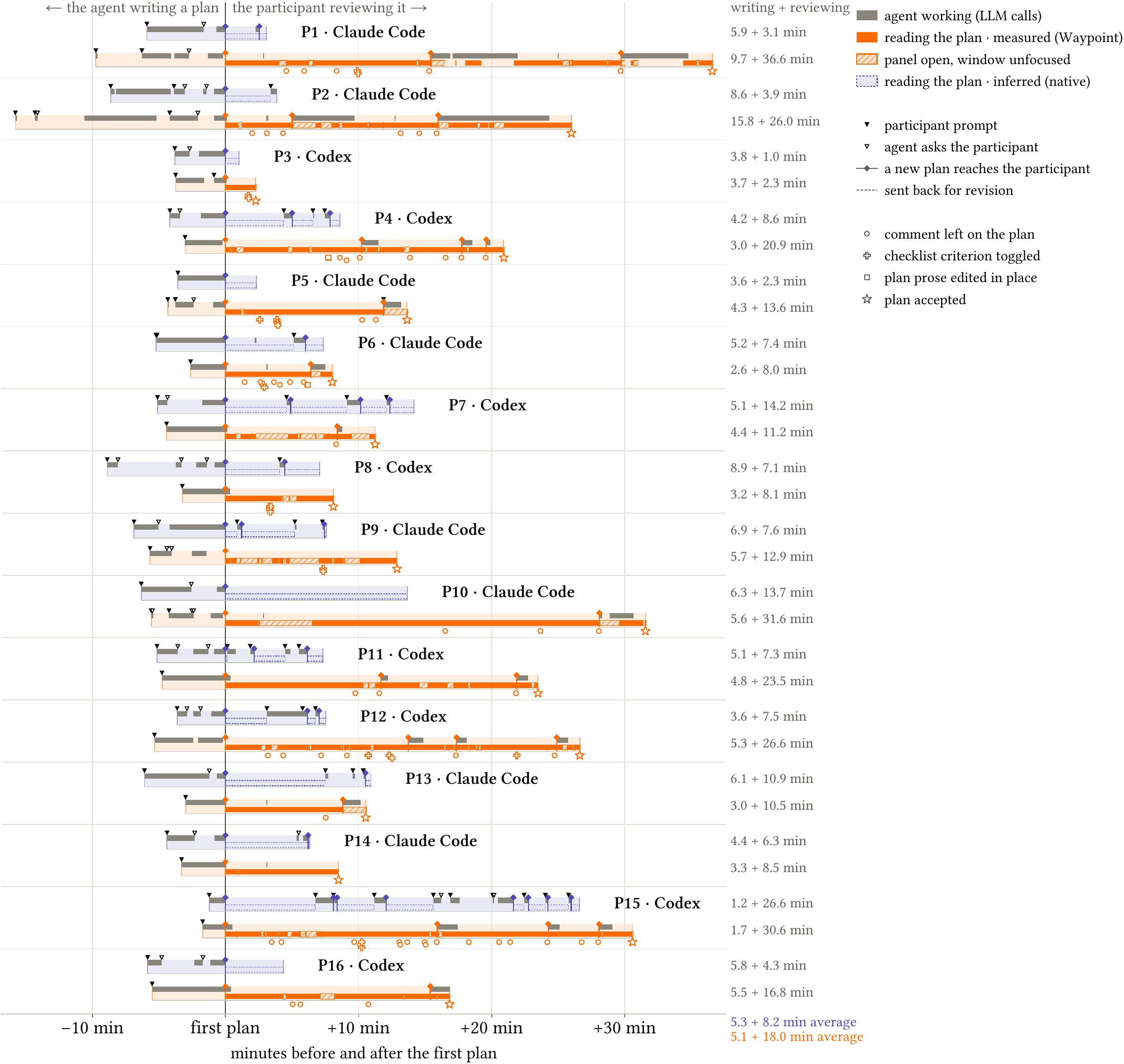}
    \caption{Planning and review timelines for all 32 sessions, aligned to the moment the first plan reached the participant (time zero). For each participant, the upper purple row shows native plan mode, and the lower orange row shows \textsc{Waypoint}. Grey segments show agent activity. Solid orange bands show intervals when the \textsc{Waypoint} plan viewer was open and focused. Outlined purple bands show native review intervals inferred from the agent presenting a plan and the participant's next prompt. Symbols mark participant and agent turns, plan submissions, revision requests, comments, checklist and prose edits, and plan acceptance.}
    \Description{Paired horizontal timelines show planning activity for participants P1 to P16, labelled with their assigned coding agent. Time runs left to right in minutes, with ticks ten minutes apart and a vertical line at the first plan's arrival. Initial writing appears to the left of zero, with review and further agent activity to the right. Each pair places native plan mode above Waypoint. Hatched Waypoint intervals indicate that the viewer remained open while its window was unfocused. Event markers cluster differently across sessions, with many comments during P15's Waypoint review and repeated native plan submissions for P15. The right margin lists initial writing and subsequent review-and-revision durations separately. Waypoint review-and-revision durations range from 2.3 minutes for P3 to 36.6 for P1, and native durations from 1.0 for P3 to 26.6 for P15.}
    \label{fig:rq1_review}
\end{figure*}

We provide an overview of all 32 planning sessions, aligned with the moment when the first plan reached the participant, as shown in Figure~\ref{fig:rq1_review}. First-plan generation took a similar time in both conditions. The subsequent review and revision loop was significantly longer with \textsc{Waypoint} ($M=18.0$ minutes, $SD=10.2$) than with native plan mode ($M=8.2$, $SD=6.1$; $W=7$, $p=.001$). Fourteen participants spent more time in the \textsc{Waypoint} loop, and two in native plan mode. This interval includes participant activity and agent revision time. Across the 16 sessions in each condition, agents submitted 37 \textsc{Waypoint} and 39 native plan iterations. Only \textsc{Waypoint} persisted these submissions as explicit versioned artefacts, so we used participant-initiated revision rounds as the cross-condition proxy for reviewer-requested plan changes. Participants initiated 21 revision rounds with \textsc{Waypoint} and 21 with native plan mode. Native agents made two additional self-corrections without a new participant request. The longer \textsc{Waypoint} loop therefore did not reflect more participant-initiated revision cycles.

Across all sessions, participants made 99 distinct review interventions. 67 \textsc{Waypoint} interventions consisting of 56 comments, eight acceptance-gate changes, two direct plan edits, and one cross-cutting request. 32 native interventions consisted of 29 prompts and three plan rejections. Some of these contained several separate requests bundled into one prompt. We therefore split these interventions into individual requests, questions, changes, or endorsements. This unbundling prevents a prompt containing multiple requests from being counted the same as a comment containing only one. This produced 137 coded interventions, with 83 in \textsc{Waypoint} and 54 in native plan mode. 

We coded each intervention by what it addressed. Counts are reported as \textsc{Waypoint} \textit{vs.} native plan mode. Participants requested tests, acceptance gates, or supporting evidence (20 \textit{vs.} 6). They corrected or added required software behaviour (15 \textit{vs.} 7). They constrained implementation choices (6 \textit{vs.} 7). Participants asked for clarification (17 \textit{vs.} 11) or requested further detail about existing plan content (8 \textit{vs.} 3). They requested changes to the plan's presentation (7 \textit{vs.} 12) or steered the agent and session process (2 \textit{vs.} 8). Participants also explicitly endorsed a plan section or proposal (8 \textit{vs.} 0).

We then examined whether these interventions were reflected in the accepted plans. Of the 137 coded interventions, 111 were reflected (73 \textsc{Waypoint}, 38 native). Another 14 were clarified without changing the plan (8 \textsc{Waypoint}, 6 native). Two native requests were not reflected. Ten interventions concerned session steering rather than plan content and were coded as not applicable (2 \textsc{Waypoint}, 8 native).

Review nevertheless remained selective in both conditions. Seven sessions contained no coded intervention. However, \textsc{Waypoint} gave participants a direct way to address the parts they chose to review. P11 explained: \emph{``when I wanted to comment something, I didn't have to go and type in the chat and copy paste the part [...] we did that in the plan itself, so that was what helped me the most''}. \textsc{Waypoint} may have helped participants turn selective attention into focused, actionable feedback. After both sessions, 15 of 16 participants selected \textsc{Waypoint} as the workflow that made it easier to give feedback on the plan.

\subsection{RQ2: Understanding and Authorisation}\label{sec:results-rq2}

\begin{figure*}[t]
\centering
\includegraphics[width=\linewidth]{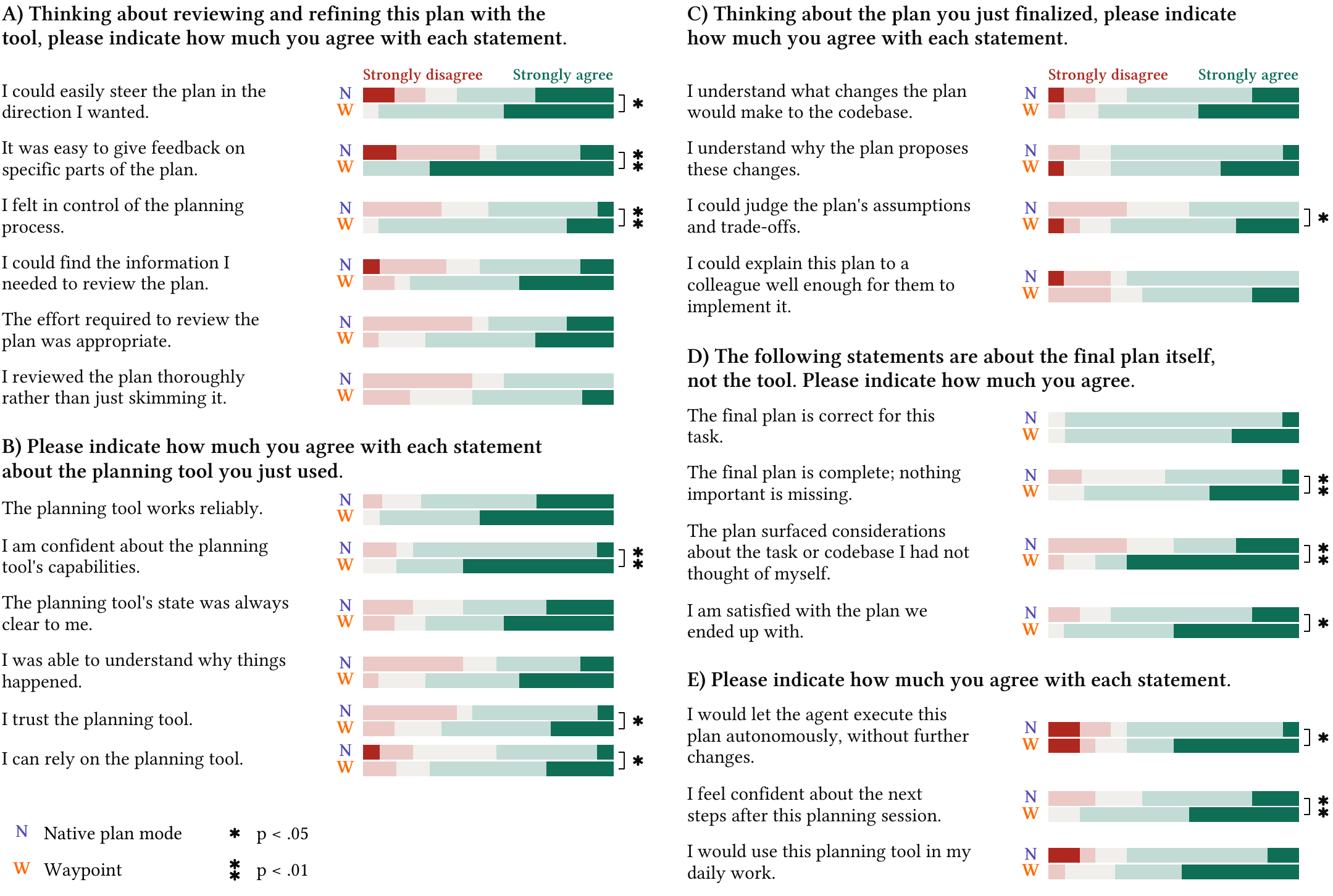}
\caption{Post-task responses under native plan mode (N) and \textsc{Waypoint} (W). Participants rated each statement on a five-level Likert scale from strongly disagree to strongly agree. Panels show (A) review and steering, (B) trust in the planning workflow, (C) self-reported understanding, (D) perceived plan quality, and (E) readiness and intended use. Stacked bars show the share of responses in each category. Asterisks mark significant paired Wilcoxon comparisons with $p<.05$ and $p<.01$.}
\Description{Paired horizontal stacked bars compare responses to 23 questionnaire items across five panels. Panels A and B are on the left, with C, D, and E on the right. Each item has an upper bar labelled N for native plan mode and a lower bar labelled W for Waypoint. Participants rated each statement on a five-level Likert scale from strongly disagree to strongly agree. Each bar represents all responses to that item in that condition. Segments run from strongly disagree on the left through disagree, neutral, and agree to strongly agree on the right. Segment widths encode response shares, without numeric axis ticks or segment values. All Waypoint responses about giving specific feedback fall in the two agreement categories. Perceived plan correctness attracts almost entirely agreement in both conditions. Brackets connect paired bars for the comparisons marked with asterisks.}
\label{fig:rq2}
\end{figure*}

The more active and localised review observed in RQ1 was accompanied by significantly higher ratings on several review and plan-quality items (Figure~\ref{fig:rq2}A,C,D). Participants found it easier to give feedback on specific parts of the plan with \textsc{Waypoint} ($M=4.71$ \textit{vs.} $M=2.86$, $W=0$, $p=.002$, $n=14$) and reported greater control ($M=4.13$ \textit{vs.} $M=3.25$, $W=0$, $p=.004$). They also felt better able to judge the plan's assumptions ($M=3.81$ \textit{vs.} $M=3.13$, $W=3$, $p=.031$) and more confident about the next steps ($M=4.31$ \textit{vs.} $M=3.63$, $W=0$, $p=.005$). Participants rated \textsc{Waypoint} plans as more complete, more likely to surface considerations they had missed, and more satisfying ($p\leq.013$ for each). Perceived correctness did not differ significantly, however, P5 said the presentation helped them \emph{``challenge it a bit better''}. Neither did any of the six NASA-TLX dimensions. We expected this because \textsc{Waypoint} was designed to create opportunities for deeper review. Participants could choose to spend more time and intervene in more detail when the added structure was useful, which may explain why perceived workload did not decrease.
Each \textsc{Waypoint} feature group received a median rating of ``Very useful'' (4), with nine to eleven participants assigning the highest rating. We asked for ratings on assumption callouts, decision callouts, acceptance checklists, detail on demand with file references, comments/edits/rejections, and the plan as a versioned file.

The greater sense of control extended to participants' readiness to authorise the plan (Figure~\ref{fig:rq2}E). Participants were significantly more willing to let the agent execute the \textsc{Waypoint} plan autonomously ($M=3.88$ \textit{vs} $M=3.38$, $W=4$, $p=.035$). After completing both conditions, 14 of 16 also selected \textsc{Waypoint} as making them more confident about large or risky changes. Together with their greater confidence about the next steps, these results suggest that participants felt more ready to proceed with the \textsc{Waypoint} plan. These measures capture stated readiness in the study rather than authorisation decisions in participants' own projects.

Participants' reports of understanding varied across measures. The three items asking what the plan would change, why it proposed those changes, and whether participants could explain it to a colleague, and the comprehension quiz showed no significant difference (Figure~\ref{fig:rq2}C). In the direct comparison, however, 15 of 16 participants selected \textsc{Waypoint} as giving them a better understanding of what the agent would do.

The trust measures reflected this greater readiness to proceed. Participants rated \textsc{Waypoint} significantly higher for confidence in the planning workflow's capabilities ($M=4.43$ \textit{vs.} $M=3.64$, $W=0$, $p=.004$, $n=14$), trust ($M=3.81$ \textit{vs.} $M=3.25$, $W=10$, $p=.029$), and reliance ($M=3.87$ \textit{vs.} $M=3.27$, $W=0$, $p=.014$, $n=15$). The reliability item showed similar responses across conditions (Figure~\ref{fig:rq2}B). Interview responses illustrate further how task scope could shape this preference. P8 generally skipped plan mode for small features but would \emph{``clearly use \textsc{Waypoint}''} for \emph{``large features that touch many files''}. Together, these findings suggest that \textsc{Waypoint} supported greater stated readiness to authorise agent execution, especially for larger changes where participants wanted a structured basis for review.

\subsection{RQ3: Accepted Commitments Remained Traceable Through Autonomous Implementation}\label{sec:results-rq3}

After participants accepted their plans, we followed all 32 plans through their corresponding autonomous implementation workflows. Participants did not inspect the resulting code or records. All 32 planned implementations and the 16 no-plan references passed every ticket-valid test. Each Task~1 implementation passed all 15 tests, and each Task~2 implementation passed all 13 tests. These uniform results show that every implementation satisfied the tested acceptance behaviour, while leaving differences in plan coverage, implementation completeness, and verification outside the scope of the ticket tests. For example, P1's native implementation passed every ticket-valid test despite stopping after the backend work, omitting the planned web and mobile changes, and failing type checking. We therefore examined how accepted commitments reached implementation and considered additional measures beyond the ticket tests.

RQ1 identified 137 unbundled review requests. For RQ3, we followed the 51 requests that entered the accepted plans and were assessed as implementable commitments. These comprised 22 corrections to required software behaviour, 18 requests for tests, gates, or supporting evidence, and 11 implementation choices. The commitments came from 13 participants across 18 planning sessions. We coded a commitment as followed when the code, tests, or recorded agent actions showed that it had been implemented. We coded it as not followed when the retained evidence showed that it was missing or incomplete. Overall, 42 commitments were followed, and nine were not. The \textsc{Waypoint} plans contained 33 commitments, of which 31 were followed, and two were not. Native plans contained 18 commitments, of which 11 were followed, and seven were not. Among the seven native commitments not followed, six concerned planned tests, final verification, or incremental delivery that did not occur. The remaining commitment concerned an accepted data-access refactoring: the implementation created the repository module but routed only one call site through it and filtered the remaining call sites inline. Of the two \textsc{Waypoint} commitments not followed, one implemented the server behaviour but omitted the requested restore interface. The other violated an accepted implementation constraint by seeding the list's sort order in a post-hydration effect that the plan had explicitly ruled out. Table~\ref{tab:rq3_intent_examples} illustrates examples of followed and not-followed commitments.

\begin{table*}[t]
\caption{Examples of participant-introduced commitments traced from review to the accepted plan and implementation.}
\label{tab:rq3_intent_examples}
\Description{Four-column table tracing five selected commitments across review, planning, and implementation. Each row identifies the participant, workflow, and adherence classification, then presents the participant's request, the accepted plan, and the observed outcome from left to right. The first two rows show followed commitments concerning Trash retention and search indexes. The remaining three show an omitted requirement audit, an incomplete data-access refactoring, and a sort-order implementation that violated the accepted constraint.}
\centering
\small
\begin{tabularx}{\textwidth}{@{}p{0.13\textwidth}XXX@{}}
\toprule
    Case & Participant's commitment & Accepted plan & Implementation outcome \\
    \midrule
    P6, \textsc{Waypoint}, followed &
    Use a 30-day Trash period and show how many days remain. &
    Each item in Trash shows the time remaining before it is automatically deleted. &
    The Trash interface displayed ``auto-deletes in N days'' using the shared 30-day retention value. \\
    P7, native, followed &
    Retain keyword and vector indexes while keeping trashed bookmarks out of search. &
    Soft deletion updates indexed \texttt{deletedAt} metadata instead of removing the existing entries. &
    Keyword and vector searches filtered on \texttt{deletedAt}, while restoration reused the retained index entries. \\
    P8, native, not followed &
    Embed the ticket verbatim so the implementation can be checked against every requirement afterwards. &
    A final ``Ticket Feedback Loop'' compares the result with each requirement, the out-of-scope list, and the acceptance criteria, recording any unmet item. &
    The session log contains the feedback loop only in the echoed plan. No requirement-by-requirement audit was performed before the run ended. \\
    P15, native, not followed &
    Refactor the implementation to avoid changing 83 call sites individually. &
    Phase 1 introduces a shared \texttt{BookmarksRepo} data-access layer. &
    The implementation added the repository module but routed only one call site through it. The other call sites gained inline \texttt{deletedAt} filters across 34 files. \\
    P5, \textsc{Waypoint}, not followed &
    Decide the list's sort order fully server-side. &
    The list page computes \texttt{isManualList} and \texttt{initialSortOrder} at render time and seeds the store on initial render, not via a \texttt{useEffect} that runs after hydration. &
    The page passed both values as props, but the grid seeded the store in a post-hydration \texttt{useEffect}, the mechanism the plan had excluded. \\
    \bottomrule
\end{tabularx}

\end{table*}

The workflows also left different verification material (Table~\ref{tab:rq3_implementation}). Every \textsc{Waypoint} implementation added at least one test case, compared with 13 of 16 native implementations. \textsc{Waypoint} implementations added a mean of 7.6 test cases, compared with 3.3 in the native workflow. Two native implementations each introduced one newly failing repository test, while none of the \textsc{Waypoint} implementations did. Type checking passed for 15 of 16 \textsc{Waypoint} implementations and 13 of 16 native implementations.

\begin{table*}[t]
\caption{Descriptive verification and cost outcomes for the 16 native and 16 \textsc{Waypoint} implementation runs.}
\label{tab:rq3_implementation}
    \Description{Seven-column table with one row for native execution and one for Waypoint execution. After the workflow name, columns report runs adding test cases, mean test cases added, runs without newly failing tests, runs passing type checking, mean execution time in minutes, and mean cost in United States dollars. Run counts use a denominator of 16 in each workflow. Test-case counts, time, and cost are means per run.}
\centering
\small
\begin{tabular}{@{}lcccccc@{}}
\toprule
Workflow &
\shortstack{Runs adding\\test cases} &
\shortstack{Mean test\\cases added} &
\shortstack{No newly\\failing tests} &
\shortstack{Type check\\passed} &
\shortstack{Mean\\minutes} &
\shortstack{Mean\\cost} \\
\midrule
Native & 13/16 & 3.3 & 14/16 & 13/16 & 15.0 & US\$3.29 \\
\textsc{Waypoint} & 16/16 & 7.6 & 16/16 & 15/16 & 21.7 & US\$4.15 \\
\bottomrule
\end{tabular}
\end{table*}

The \textsc{Waypoint} workflow also retained criterion-level completion records. Its 16 plans contained 210 criteria, of which 161 agent-checkable criteria were marked complete with dedicated evidence records, while 49 required human judgement and remained unevaluated. P6 described the intended value of such evidence as being able to inspect a verification mark and see \emph{``with a justification, in what way it thinks it verified that''}. Autonomous implementation also took longer and cost more in the \textsc{Waypoint} workflow, with means of 21.7 minutes and US\$4.15 for \textsc{Waypoint}, compared with 15.0 minutes and US\$3.29 for native plan mode. Together, the accepted plans and generated evidence provided a reference for assessing whether participants' review decisions reached implementation, including ticket-level differences that tests did not reveal.

\section{Discussion}

An implementation plan is the final proposal a developer can review before an agent modifies the codebase. Here, we examine what our results show about that moment. What acceptance turns the plan into, how a reviewable format directs a developer's limited attention and its associated costs, and what is substantively granted through acceptance.

\subsection{Plans as Authorisation Artefacts}

Before a developer accepts, a plan is the agent's forecast of what it intends to do. After acceptance, it becomes a record of what the developer allowed to proceed. In our study, these records often contained more than the ticket requested. Participants added behavioural requirements (e.g., a 30-day retention period for the Trash), constrained implementation choices, or asked for tests and evidence. These interventions were then transferred into the code. Of the assessable commitments, 31 of 33 in \textsc{Waypoint} plans and 11 of 18 in native plans were followed into implementation. 
Situated accounts of planning clarify what such a record can and cannot do. A plan is a resource that people and agents interpret in action, not a script that determines it~\citep{1212951,suchmanResponseVeraSimons1993}. Plans should therefore be referenced alongside their implementation records.

For the record to support a later review, it should show how the agent implemented each part of the task, where it deviated, and what was excluded or not checked. P1's native implementation illustrates this: it passed all ticket-valid tests but omitted the planned web and mobile changes and failed type checking, so the simple coverage tests would have called it complete. P6 wanted each verification gate to explain \emph{``with a justification, in what way it thinks it verified that''}. We summarise three properties necessary for plans to serve as an authorisation artefact. Each commitment should retain its origin and a stable identity. Completion records should link to the commitment, provide evidence of execution, or identify where implementation did not follow it. The record distinguishes the agent's account of its own work from an independent check. This allows the agent's output to be audited rather than taken on the agent's own word~\citep{gordonCoAuditTools2023}.

\subsection{Designing for Selective Review and Paying for It}

Participants reviewed selectively in both conditions, as our formative interviews had indicated. What changed with \textsc{Waypoint} was the output of their limited attention. Interventions shifted toward verification (20 versus 6) and required behaviour (15 versus 7) and away from presentation (7 versus 12) and session steering (2 versus 8). In other words, participants in the native condition requested more on getting the plan into a reviewable shape, whereas with \textsc{Waypoint} the format had already done that work. Participants also preferred to steer with the agent rather than author the plans directly. They left 56 comments and made only two direct changes. P11 put it plainly, as with \textsc{Waypoint}, \emph{``I didn't have to go and type in the chat and copy-paste the part[s]''}. Therefore, plan content should provide addressable anchors so that a question, a correction, or a request for evidence attaches to the assumption or decision it concerns rather than to the plan as a whole. Repetitive organisational work should be handled by the format, leaving reviewers free to focus on behaviour and verification. Furthermore, the interface should assume selective reading and be designed for it, using progressive disclosure and visible assumptions to direct attention to consequential choices. Code-review research shows that redirecting limited attention can improve reviewers' awareness of risky changes or add helpful revision comments~\citep{yuStudyingImpactRisk2024,olewickiEmpiricalStudyCodeReview2024}. Selective review is, therefore, a design target with a failure mode of its own.

Nevertheless, this review-focused planning process adds additional cost. The \textsc{Waypoint} review and revision loop averaged 18.0 minutes, compared to 8.2 minutes for the native plan mode. Autonomous implementation took 21.7 minutes and \$4.15 per run, versus 15.0 minutes and \$3.29. The longer review interval accompanied more interventions (83 versus 54), a shift toward behaviour and verification, and higher reported control and readiness. During implementation, \textsc{Waypoint}, however, added more test cases, introduced no new failing repository tests, and returned completion records for all agent-checkable criteria. These time and cost differences concern the complete workflows. The \textsc{Waypoint} workflow produced more review requests and test cases and retained dedicated criterion-level completion records.

\subsection{Authorisation as a Bounded, Risk-Scaled Handoff}

Accepting a plan grants permission for a bounded proposal. It does not certify the plan or predict the code. Conditional delegation follows a similar mechanism by allowing people to mark the regions where an AI may act on its own while leaving the rest to human judgement~\citep{laiHumanAICollaborationConditional2022}. Participants seemed to manage both by scaling their authorisation to the task at hand. Fourteen of 16 preferred \textsc{Waypoint} for large or risky changes, and P16 would reach for it on a new feature but switch to the faster native workflow for a small bug fix. There is a need for varying levels of coding-agent oversight across tasks and risk levels~\citep{shuklaHedwigDynamicAutonomy2026}. Distinguishing between acceleration, when a programmer knows what comes next, and exploration, when they do not, offers a way to think about when which workflow is worth its price~\citep{barkeGroundedCopilot2023}.

The harder question is why participants felt ready to authorise the agent's plan. Participants reported being more willing to let the agent execute a \textsc{Waypoint} plan. The comprehension quiz, which covered repository knowledge and understanding of specific implementation details in the plan, did not differ significantly between conditions. Users may not need to understand the entire codebase, but they should be able to judge the scope, assumptions, and evidence of the specific plan they are approving. Authorisation interfaces should keep these visible, allow users to revise them, and require renewed approval when they change. We refer to this as justified delegation: authorising a particular proposal on the basis of its stated terms and supporting evidence, rather than relying on general trust in the agent. Planning cannot remove the uncertainty of autonomous coding, but it can make that uncertainty open to challenge before authority passes to the agent and preserve a record of what was approved.

\section{Limitations and Future Work}

We evaluated \textsc{Waypoint} as a complete workflow against native plan mode. The conditions differed in plan generation, presentation, review, authorisation, and execution support. Our findings therefore concern the bundled workflows rather than individual affordances.
Our user study has several limitations. Participants spent relatively little time in an unfamiliar production repository. This limited the codebase knowledge they could develop before the comprehension quiz. To make complete plan review feasible, we selected tasks that crossed several components but fit within a single session. Further, the study covered only two tasks from one web application, which may limit the ecological validity of our findings beyond this repository and task setting. Our sample was also weighted toward developers with five or fewer years of professional experience. Their review practices may not reflect those of more experienced practitioners.
Longitudinal deployments could examine planning and review with developers working in their own repositories. This setting would also support larger and more complex tasks. Participants did not inspect the execution records generated after they accepted their plans. Future studies could examine how accepted plans, execution records, and resulting implementations help developers to understand what the agent produced and identify deviations from the plan. Finally, our evaluation studied plan review and authorisation between a single participant and a single primary agent. Software engineering instead distributes knowledge and authority across teams and different roles. Future work should explore collaborative plan review as an (N{:}N) relationship among engineers, other stakeholders, and agents. Plans could record who proposed, revised, and approved reviews or route decisions to specialists.

\section{Conclusion}

As coding agents become capable of implementing over long horizons, overseeing what work is done and how it is done becomes a design problem in its own right. We investigate implementation plans as the boundary between a developer's request and autonomous execution. Drawing on formative interviews with practitioners, we designed requirements for review and revision, built \textsc{Waypoint}, compared it with native planning, and implemented each accepted plan. \textsc{Waypoint} exposes assumptions, decisions, and acceptance criteria for review before authorisation, then preserves the accepted plan alongside returned evidence. Our findings present reviewable plans as authorisation artefacts that make planned work contestable before execution and provide a reference for examining implementations afterwards.

\bibliographystyle{ACM-Reference-Format}
\bibliography{bibliography}

\clearpage
\appendix

\section{Participants}\label{sec:participants-table}

Table~\ref{tab:participants} lists the self-reported background of the sixteen participants.

\begin{table}[h]
    \caption{Participant background. Experience is years of professional or academic software development. AI use is the reported frequency of using AI coding assistants, and Plan mode the reported frequency of using a dedicated plan mode before letting an agent change code.}
    \label{tab:participants}
    \Description{Five-column table with one row per participant, P1 to P16. Columns give years of development experience, self-described role, frequency of AI coding assistant use, and frequency of plan-mode use. Roles span students, junior, mid-level, and senior developers, a tech lead, and a researcher. Experience ranges from less than one year to five to ten years. Fourteen participants use AI assistants daily and two weekly. Plan-mode use ranges from never to always.}
    \centering
    \small
    \begin{tabular}{@{}lllll@{}}
        \toprule
        ID & Experience & Role & AI use & Plan mode \\
        \midrule
        P1 & 3--5 years & Student & Daily & Often \\
        P2 & 5--10 years & Senior Developer & Daily & Rarely \\
        P3 & 3--5 years & Tech lead & Daily & Rarely \\
        P4 & $<$1 year & Student & Daily & Always \\
        P5 & $<$1 year & Student & Weekly & Rarely \\
        P6 & 1--3 years & Researcher & Daily & Never \\
        P7 & 1--3 years & Junior Developer & Daily & Often \\
        P8 & 1--3 years & Junior Developer & Daily & Sometimes \\
        P9 & 5--10 years & Senior Developer & Daily & Often \\
        P10 & 1--3 years & Student & Daily & Often \\
        P11 & 3--5 years & Mid-Level Developer & Daily & Often \\
        P12 & 3--5 years & Junior Developer & Daily & Often \\
        P13 & 3--5 years & Junior Developer & Daily & Sometimes \\
        P14 & 3--5 years & Mid-Level Developer & Daily & Rarely \\
        P15 & 1--3 years & Junior Developer & Daily & Often \\
        P16 & 3--5 years & Researcher & Weekly & Never \\
        \bottomrule
    \end{tabular}
\end{table}

\section{Study Tasks}\label{sec:study-tasks}

We presented both tasks as product backlog tickets with background, requirements, exclusions, and an acceptance statement.

\subsection{Task 1: Recoverable Delete}

The first ticket described accidental deletion as a risk for people who use Karakeep as long-term memory. It asked participants to plan a Trash feature with the following requirements:

\begin{enumerate}[leftmargin=*]
    \item Deleting a bookmark moves it to Trash rather than removing it immediately.
    \item A user can restore a bookmark with its lists, tags, and favourite state intact.
    \item A user can permanently delete a bookmark from Trash.
    \item Trashed bookmarks do not appear in the bookmark grid, tag or list views, or search.
    \item A Trash view provides restore and permanent-delete actions, with a quick undo after deletion.
    \item Automatic cleanup of old trashed bookmarks is optional.
\end{enumerate}

Trash for other entities, per-list Trash, shared-list behaviour, and a user-configurable cleanup interval were out of scope. The acceptance statement required that a bookmark remain recoverable and intact until permanent deletion while remaining absent from normal browsing and search.

\subsection{Task 2: Manual List Ordering}

The second ticket asked participants to plan how people could arrange bookmarks within a list:

\begin{enumerate}[leftmargin=*]
    \item A user can reorder bookmarks within a list, and the order persists.
    \item A user can pin a bookmark to the top of a list.
    \item The arrangement remains after reload and on another device.
    \item Ordering is independent per list, so the same bookmark can occupy different positions in different lists.
\end{enumerate}

Global bookmark ordering, ordering lists or tags, and manual ordering of smart lists were out of scope. The acceptance statement required saved order and pinning to survive reload without affecting another list.

\section{Coding and Derived Variables}\label{sec:coding-audit}

We began by turning the interaction records into comparable units. An \emph{act} was one typed turn in native plan mode or one located \textsc{Waypoint} interaction. We counted acts directly from the records. Within each act, every request, question, change, or endorsement that could be answered or satisfied independently counted as one \emph{ask}. We applied this rule in both conditions, allowing a native message containing several requests to be compared with individual \textsc{Waypoint} interactions. Each remaining ask then received one purpose code. Table~\ref{tab:coding-overview} presents the four coding steps in this order.

\begin{table*}[t]
    \caption{Codebook for review interventions and commitment tracing. Rows follow the order of the analysis, and purpose codes appear in precedence order.}
    \label{tab:coding-overview}
    \Description{Three-column table mapping four coding steps to their units and category definitions. Read each row from Coding to Unit to Values and definitions. Ask count applies to one interaction act. Purpose and plan outcome each apply to one independently answerable ask. Implementation adherence applies to one traced commitment. The purpose row contains eight categories, the plan-outcome row five, and the adherence row three. The table distinguishes a change entering the accepted plan from that commitment being followed during implementation.}
    \centering
    \small
    \begin{tabularx}{\textwidth}{@{}>{\raggedright\arraybackslash}p{0.15\textwidth}>{\raggedright\arraybackslash}p{0.10\textwidth}X@{}}
        \toprule
        Coding & Unit & Values and definitions \\
        \midrule
        Ask count & One act &
        The number of independently answerable or satisfiable requests in the act. A pasted ticket remained one ask, whereas separately satisfiable requirements were split. A rationale for a request and an approval preceding it did not count as additional asks. \\
        \addlinespace
        Purpose & One ask &
        \emph{Explicit approval} endorses the referenced plan content. \emph{Session steering} targets the agent process, tooling, or active plan version. \emph{Tests, gates, or evidence} requests verification. \emph{Behaviour correction} changes what the software should do. \emph{Implementation choice} constrains how it should be built without changing user-visible behaviour. \emph{Clarification} asks a determinate question. \emph{Further detail} expands existing plan content. \emph{Presentation} changes the plan's structure or form. \\
        \addlinespace
        Plan outcome & One ask &
        \emph{Reflected} means that the accepted plan contains the requested change. \emph{Partly reflected} means that it contains only part or a weakened form. \emph{Not reflected} means that the change is absent. \emph{Clarified without change} means that a question was resolved without changing the plan. \emph{Not applicable} is reserved for session-steering asks that do not target plan content. Partly reflected was defined but not observed. \\
        \addlinespace
        Implementation adherence & One traced commitment &
        \emph{Followed} means that retained code, tests, or recorded agent actions satisfy the accepted commitment. \emph{Not followed} means that the implementation omits, contradicts, or only partly fulfils the commitment such that its intended consequence is not met. \emph{Not assessable} means that the retained artefacts cannot establish an outcome. \\
        \bottomrule
    \end{tabularx}
\end{table*}

Next, we compared each \emph{ask} with the accepted plan and coded its plan outcome. A request became a traced commitment when it concerned required behaviour, an implementation choice, or tests and evidence, and the accepted plan reflected it. Thereafter, we followed the commitment into the autonomous run. Comparing its wording with the files the agent changed or created, and checked the execution log and regression results. This record determined whether the agent followed the commitment, departed from it, or left too little evidence to decide. We excluded human-only gates because no reviewer was present during implementation. We also excluded three requests that changed only the plan's wording or organisation and criteria written by the agent rather than the participant.

Our purpose codebook began with a close reading of the interventions. The initial scheme grouped together changes to the plan's presentation, instructions about how the agent should work, and approvals of particular proposals. We separated these actions into the eight purpose codes in Table~\ref{tab:coding-overview}. Each author then independently assigned purpose and plan-outcome codes to the asks and adherence codes to the implementation traces. When their labels differed, they returned to the participant's words and the corresponding plan or run record, discussed the case, and agreed on a final code.

\section{Study Snapshot}\label{sec:study-snapshot}

Participants used Claude Code \texttt{2.1.211} with \texttt{anthropic/\allowbreak claude-sonnet-5} and Codex CLI \texttt{0.147.0} with \texttt{gpt-5.6-sol}.

\section{Availability}\label{sec}

\textsc{Waypoint} is publicly available as the \emph{Waypoint Plan Reviewer} extension on the Visual Studio Code Marketplace (\url{https://marketplace.visualstudio.com/items?itemName=LudwigFelder.waypoint-plan-reviewer}) and can be installed with \texttt{code --install-extension LudwigFelder.waypoint-plan-reviewer}. Its source code and agent skills are available at \url{https://github.com/ludwigfelder/waypoint_extension}.

\end{document}